\documentclass[onecolumn]{cup-hpl}

\usepackage{bm}
\usepackage{siunitx}
\usepackage{booktabs}
\usepackage{subcaption}
\usepackage{hyperref}
\hypersetup{
  hidelinks,
  pdftitle={Multi-Objective Design of C-Shaped Targets for Energy--Divergence Control in Laser-Driven Proton Acceleration},
  pdfauthor={Mohammad Rezaei-Pandari; Mahdi Shayganmanesh; Mohammad Hossein Mahdieh},
  pdfkeywords={laser-driven proton acceleration; target-normal sheath acceleration; structured targets; multi-objective Bayesian optimization; particle-in-cell simulations}
}
\usepackage{microtype}

\newcommand{\Ecut}{E_{\mathrm{cut,H}}}
\newcommand{\DivH}{\sigma_{\theta,H}}
\newcommand{\etaH}{\eta_{\mathrm{H}}}

\title{Multi-Objective Design of C-Shaped Targets for Energy--Divergence Control in Laser-Driven Proton Acceleration}
\shorttitle{Multi-Objective Design of C-Shaped Targets}
\shortauthor{Rezaei-Pandari et al.}

\author[1,*]{Mohammad Rezaei-Pandari}
\author[1]{Mahdi Shayganmanesh}
\author[1]{Mohammad Hossein Mahdieh}
\address[1]{Department of Physics, Iran University of Science and Technology, Narmak, Tehran, Iran}
\corresp{Mohammad Rezaei-Pandari, \email{rezaei.m.p@gmail.com}}
\usepackage{lineno}
\makeatletter
\def\ps@plain{%
  \let\@oddhead\@empty\let\@evenhead\@empty
  \def\@oddfoot{\footnotesize\hfill\thepage\hfill}%
  \let\@evenfoot\@oddfoot
}
\makeatother
\begin{document}

\begin{abstract}
We optimize the energy--divergence trade-off of laser-driven protons from a C-shaped target using two-dimensional particle-in-cell simulations and constrained multi-objective Bayesian optimization. Five geometry and focusing parameters are varied to maximize proton cutoff energy and minimize the RMS divergence of forward protons with $E_k\geq\SI{5}{MeV}$ under a proton-only conversion-efficiency constraint of $\eta_H\geq9\%$. Of 82 designs, 60 are feasible and 15 Pareto-optimal. The Max-Energy and Min-Divergence endpoints yield 28.64~MeV at $52.51^\circ$ and 11.47~MeV at $11.13^\circ$, respectively; a representative compromise yields 21.44~MeV at $38.44^\circ$. Matched-stage diagnostics link these regimes to optical-field buildup, electron energization, charge separation, and outer-sheath organization. Numerical tests support the principal trends. Reflection timing does not support a monotonic trapping-duration law, and the charge--sheath relation remains non-local. These results establish a design-resolved route to controlling proton energy and angular quality.

\end{abstract}
\keywords{laser-driven proton acceleration; target-normal sheath acceleration; structured targets; multi-objective Bayesian optimization; particle-in-cell simulations}

\maketitle


\section{Introduction}
\label{sec:introduction}

Laser-driven ion acceleration provides a compact route to intense proton and ion sources for high-energy-density physics~\cite{Patel2003,Dyer2008,BaillyGrandvaux2025}, proton radiography~\cite{Cobble2002,Mackinnon2004}, fast-ignition concepts~\cite{roth2001fast}, and other applications requiring short, high-current proton bursts~\cite{Dromey2016,Prasselsperger2021}.

Depending on laser intensity, pulse duration and temporal contrast, target areal density, and preplasma evolution, several physically distinct---but potentially overlapping---acceleration regimes can be accessed. Radiation-pressure acceleration (RPA) transfers laser momentum to an overdense thin foil~\cite{Henig2009RPA}; breakout-afterburner (BOA) acceleration exploits enhanced coupling after an initially opaque target becomes relativistically transparent~\cite{Yin2011BOA}; collisionless-shock acceleration (CSA) reflects ions from an electrostatic shock in a suitably tailored plasma~\cite{Haberberger2012CSA}; and magnetic-vortex acceleration (MVA) has been investigated using electron-current-driven electromagnetic fields in near-critical-density targets~\cite{Tazes2024MVA}. Recent experiments further show that high-energy proton beams can emerge from cascaded or hybrid combinations of radiation pressure, transparency-related dynamics, sheath acceleration, and Coulomb repulsion rather than from a single isolated mechanism~\cite{Ziegler2024Cascaded,Shou2025Hybrid}.

Target-normal sheath acceleration (TNSA) remains among the most robust laser-driven ion-acceleration mechanisms: laser-heated electrons traverse and expand around the target, establishing an electrostatic sheath that accelerates ions into vacuum~\cite{roth2013ion,gizzi2017new,Passoni2013}. Because the sheath is generated by the spatial and temporal organization of energetic electrons, proton energy and angular quality depend not only on the incident laser intensity but also on how the target geometry controls optical coupling, electron transport, and charge separation. In contrast to schemes that require carefully controlled ultrathin-foil opacity, deliberately timed relativistic transparency, or tailored near-critical-density profiles, TNSA can operate across a comparatively broad range of solid-target and laser conditions, although its broad spectra and substantial divergence remain important limitations. The present study therefore addresses geometric control of a sheath-dominated solid-target interaction rather than assuming that the optimized target operates in a pure RPA, BOA, CSA, or MVA regime.

Structured targets offer several routes to modify these processes. Resonant or nanostructured surfaces can enhance laser coupling~\cite{Ceccotti2013,vallires2021enhanced}, cone and low-density structures can reshape electron generation and transport~\cite{Kluge2012,Sgattoni2012,Prencipe2021}, and curved or microstructured targets can alter ion focusing and beam formation~\cite{roth2001fast,Bailly-Grandvaux2020,Morace2022,Choudhury2024}. Recent work has increasingly emphasized that useful source optimization is intrinsically multi-parametric: advanced targets can improve control of laser-driven hadron beams~\cite{Mirani2025AdvancedTargets}, and petawatt-facility studies now assess spectral and spatial beam properties jointly rather than treating cutoff energy alone as a sufficient performance metric~\cite{Filippov2026HPLSE,Martin2024HPLSE}. Open and closed hemisphere--cone targets further show that curved inner surfaces can generate energy-selective focusing and that closure geometry can preserve or degrade angular quality~\cite{King2024PPCF}. Ultrathin-foil experiments also demonstrate narrow proton components, but their half-angle definition is not interchangeable with the centroid-referenced RMS metric used here~\cite{Bin2013POP}. These results motivate optimization strategies that retain beam quality as an explicit objective rather than as a post hoc diagnostic.

Bayesian optimization is particularly attractive for this problem because high-fidelity particle-in-cell (PIC) simulations and experiments are too costly for exhaustive exploration of strongly coupled design spaces. Multi-parameter Bayesian optimization has already been demonstrated for laser-driven ion acceleration in PIC simulations~\cite{Dolier2022NJP}, and automated experimental optimization has shown that data-efficient search can identify improved operating conditions in laser-driven ion sources~\cite{Loughran2023HPLSE}. Machine-learning approaches have likewise been used to model laser-driven ion-acceleration responses over multidimensional parameter spaces~\cite{Djordjevic2021POP}. Surrogate-assisted experimental optimization has also increased maximum proton energy by controlling a 20-dimensional laser-wavefront space~\cite{Catrix2025APL}. A recent six-parameter multi-objective TNSA preprint optimizes peak energy, peak-window charge fraction, and in-window charge for a double-layer target~\cite{Zhang2026Arxiv}; it does not optimize divergence or consider a curved cavity with an efficiency constraint. Most closely related peer-reviewed studies, however, either optimize a scalar performance measure, scan a restricted parameter subset, or characterize a fixed target concept. The coupled energy--divergence trade-off and its underlying plasma mechanism therefore remain important design questions for structured targets.

The annular-sector, or C-shaped, target considered here was motivated by our earlier fixed-geometry study of the same target class~\cite{RezaeiPandari2026Arxiv}. That comparison linked the higher proton cutoff primarily to enhanced electromagnetic coupling and a hotter energetic-electron population, while the curved geometry reorganized the surface-resolved and angular ion populations; it did not isolate an additional cutoff-energy gain caused solely by geometric convergence. The result established the viability of the target concept but did not determine how the cavity and laser focus should be chosen when high proton energy competes with low divergence. Nor does it justify assuming that a single scalar such as optical trapping duration controls performance throughout the design space. Once the inner radius, opening angle, wall thickness, laser waist, and focal position are allowed to vary, the geometry can simultaneously modify local optical-field buildup, electron-energy deposition, charge-separation morphology, and the angular support of the accelerating sheath.

The proposed open, curved geometry is also supported by existing target-fabrication capabilities. Experiments have used coated hollow microspheres with laser-micromachined openings as staged proton-acceleration targets~\cite{Burza2011Microspheres}, while three-dimensionally printed microtubes and other engineered microstructures have already been fielded in high-intensity laser-ion experiments~\cite{Bailly-Grandvaux2020}. The preparation, patterning, and mounting of structured free-standing foils have likewise been demonstrated~\cite{Gheorghiu2021Fabrication}, and recent experiments have generated energetic proton beams from micrometre-scale structures produced by two-photon polymerization/direct laser writing~\cite{Tochitsky2025Microstructures}. Together, these results suggest practical routes to an annular-sector target, including opening or sectioning a curved micro-shell and directly writing an open annular scaffold followed, where required, by material deposition and mounting. These fabrication precedents do not imply that the exact optimized C-shaped design has already been manufactured or experimentally validated: achievable wall thickness, edge quality, opening-angle tolerance, material or coating choice, alignment, and survival of the laser pre-pulse remain experiment-specific constraints, particularly for the thinnest walls considered here.

To the best of our knowledge, no previous report has combined a C-shaped annular-sector target, simultaneous proton-cutoff-energy and centroid-referenced-divergence objectives, five coupled geometry/focusing variables, and a proton-efficiency feasibility constraint within one Bayesian PIC campaign. This qualified statement concerns that specific combination and is not a claim to the first structured-target optimization or the first Bayesian optimization of laser-driven ions.

Here we formulate this problem as a constrained multi-objective Bayesian optimization coupled directly to two-dimensional Smilei simulations. The five design variables are the target inner radius $R_{\rm in}$, laser-facing opening angle $\theta_{\rm open}$, wall thickness $d_{\rm wall}$, laser waist $w_0$, and signed focal-position offset $\Delta x_f$. The objectives are to maximize the proton cutoff energy and minimize the RMS divergence of forward-going protons with reconstructed $E_k\geq\SI{5}{MeV}$, while enforcing a proton-only laser-to-proton conversion efficiency of at least 9\%. The search domain is extended only when the accumulated constrained Pareto set shows persistent boundary pressure, preventing an initially chosen box from being treated as unquestionably sufficient.

The optimization is followed by a matched-stage mechanism analysis of representative Max-Energy, compromise, and Min-Divergence regimes. Optical field and Poynting diagnostics, electron loading and phase space, dynamic carbon-interface reconstruction, species-resolved charge separation, and proton phase space are combined to determine where the Pareto trade-off originates. The resulting picture is not a monotonic ``more trapping gives more energy'' model. Instead, optimized geometry and focusing establish different early optical-coupling conditions that reorganize electrons and near-interface charge separation, thereby changing both the strength and angular support of the outer accelerating sheath. The charge-to-field step is explicitly non-local: regime-scale strength and support are robust, while no universal pointwise charge/sheath relation is claimed. This design-resolved mechanism connects the optimization outcome directly to the laser--plasma dynamics that determine proton energy and angular quality.


\section{Numerical model and analysis methods}
\label{sec:numerical_model}

\subsection{PIC configuration and C-shaped target}
The simulations are performed with the open-source Smilei particle-in-cell code~\cite{smilei2018}, version \texttt{5.1}, in two-dimensional Cartesian geometry. The production executable reported HDF5 1.10.10, Python 3.12.3, and NumPy 2.4.4 at run initialization. The laser wavelength is $\lambda_0=\SI{0.8}{\micro\meter}$ and the production grid uses $\Delta x=\Delta y=\lambda_0/32=\SI{0.025}{\micro\meter}$ over a $\SI{50}{\micro\meter}\times\SI{34}{\micro\meter}$ domain. The timestep corresponds to a CFL fraction of 0.96, electromagnetic boundaries are Silver--M\"uller, particles are removed at the box boundaries, and production runs are evolved to \SI{420}{fs}. Nine particles per cell are used for each species. Hydrogen has mass $1836m_e$ and charge $+e$; carbon has mass $12\times1836m_e$, initial charge $+3e$, atomic number $Z=6$, and tunnel ionization enabled. The equal hydrogen/carbon ion density is 26.44 in Smilei normalized units, with the initial electron density chosen for charge neutrality.

The C-shaped target is an annular sector centered at $(x_0,y_0)=(\SI{18}{\micro\meter},L_y/2)$ and parameterized by inner radius $R_{\rm in}$, the laser-facing vacuum opening angle $\theta_{\rm open}$, and wall thickness
\begin{equation}
d_{\rm wall}=R_{\rm out}-R_{\rm in}.
\end{equation}
To phenomenologically account for surface pre-expansion associated with finite laser contrast, the radial interfaces are initialized with a prescribed short-scale pre-plasma represented by hyperbolic-tangent density transitions with characteristic scale $\delta=\SI{0.1}{\micro\meter}$. This common profile provides a controlled comparison across the campaign but is not a geometry-specific hydrodynamic prediction and does not imply identical experimental pre-plasmas for different optimized geometries. The normally incident Gaussian pulse propagates along $+x$ and uses the Smilei \texttt{tgaussian} temporal profile with \texttt{fwhm}=\SI{25}{fs} and \texttt{order}=2. In the exact production Smilei source, this parameter is the FWHM of the Gaussian field envelope; the corresponding intensity-envelope FWHM is therefore $25/\sqrt{2}\simeq\SI{17.68}{fs}$. The waist $w_0$ is optimized while the nominal laser-energy scaling is held fixed through
\begin{equation}
a_0(w_0)=10\sqrt{\frac{\SI{3}{\micro\meter}}{w_0}}.
\label{eq:a0_scaling}
\end{equation}
The focal plane is
\begin{equation}
x_f=x_0+\Delta x_f,
\label{eq:focal_offset}
\end{equation}
so positive $\Delta x_f$ moves the nominal focus downstream along $+x$ relative to the shell center.

\subsection{Matched-stage interface, electron, and charge diagnostics}
The three representative Pareto regimes are re-simulated and compared at matched interaction times $\tau=t-t_{\rm ref}$. For each target, $t_{\rm ref}$ is defined as the first crossing of 5\% of the local incident laser-field envelope peak measured $\SI{0.5}{\micro\meter}$ upstream of the detected target surface. This operational interaction reference denotes low-amplitude pulse arrival, not the pulse peak or literal first contact. At each stored time, the inner and outer carbon interfaces are reconstructed from the carbon charge-density field on target-centered polar rays. The polar angle $\phi$ labels the direction of each ray relative to the target center $(x_0,y_0)$ and therefore identifies the corresponding angular position along the target surface. A case-specific detector threshold is calibrated at $\tau=0$ against the nominal $R_{\rm in}$, $R_{\rm out}$, and $d_{\rm wall}$ and then frozen for that case. The dynamic outer interface defines the vacuum-facing normal $\mathbf n_{\rm vac}$ and tangent $\mathbf t$, giving
\begin{equation}
E_n=\mathbf E\cdot\mathbf n_{\rm vac},\qquad E_t=\mathbf E\cdot\mathbf t.
\label{eq:local_sheath_components}
\end{equation}
Here $E_n$ and $E_t$ denote the electric-field components normal and tangential to the dynamically reconstructed target surface, respectively. The unit vector $\mathbf n_{\rm vac}$ points from the target surface toward the adjacent vacuum, whereas $\mathbf t$ is the corresponding local tangent. Positive $E_n$ accelerates protons away from the target surface, while $E_t$ characterizes the field component parallel to that surface.
The principal early comparison of the accelerating sheath and electron-loading distributions uses the 0.5--\SI{1.5}{\micro\meter} vacuum-side sampling band. Sheath angular support is quantified from cumulative positive-$E_n$ weight; detailed distance-band and interface-tracking assessment are provided in Supplementary Sec. S8.
\label{subsec:dynamic_sheath_method}

Electron organization is analyzed on the same dynamic interface using the stored electron spatial statistical-weight distribution, kinetic-energy-weighted spatial distribution, $x$--$p_x$ phase space, and kinetic-energy spectrum. The ratio of the kinetic-energy-weighted distribution to the statistical-weight distribution is treated only as a spatial column mean-energy moment, while the $x$--$p_x$ phase space is used only as a longitudinal transport-direction proxy. Electron/sheath correspondence is assessed through rank/linear correlations, cosine similarity, and overlap of cumulative angular-support sectors; temporally pooled values are descriptive rather than independent significance tests.
\label{subsec:electron_transport_method}

The charge-separation bridge is evaluated from the total and species-resolved charge densities. After verifying machine-precision agreement between total charge and the species sum, positive material charge is integrated only through the dynamically detected shell, whereas negative charge is integrated over the distinct 0.05--\SI{0.50}{\micro\meter} near-interface vacuum layer. This charge-integration layer characterizes the source region immediately adjacent to the target and is distinct from the 0.5--\SI{1.5}{\micro\meter} band used to evaluate the accelerating sheath and electron loading:
\begin{align}
\Sigma_{+,\mathrm{mat}}(\phi)&=\int_{R_{\rm in}(\phi)}^{R_{\rm out}(\phi)}\max[\rho,0]\,\mathrm{d}d,\\
\Sigma_{-,\mathrm{vac}}(\phi)&=\int_{0.05\,\mu\mathrm m}^{0.50\,\mu\mathrm m}\max[-\rho,0]\,\mathrm{d}d,\\
S_{\rm pair}(\phi)&=\sqrt{\Sigma_{+,\mathrm{mat}}(\phi)\Sigma_{-,\mathrm{vac}}(\phi)}.
\label{eq:charge_pair_proxy}
\end{align}
A fixed $3^\circ$ Gaussian smoothing is used only for a common spatial-correlation assessment and is never tuned by case or time. Because the electrostatic field is non-local, $S_{\rm pair}(\phi)$ is not assumed to determine $E_n(\phi)$ pointwise. Late Max-Energy interface metrics near $\tau\sim\SI{300}{fs}$ are excluded after the independently identified tracker branch switch. Full interface, electron, charge, and thin-wall sensitivity procedures are documented in Supplementary Secs. S8--S12.
\label{subsec:charge_separation_method}

\subsection{Optical coupling and electromagnetic energy transport}
\label{subsec:optical_coupling_method}
Optical coupling is evaluated at matched interaction times using Poynting-flux probes arranged along a common rectangular integration boundary and the centerline electric-field diagnostic $E_y(x,t)$. The incident electromagnetic energy $U_{\rm inc}$ and the outgoing backward, forward, and transverse energies $U_{\rm back}$, $U_{\rm forward}$, and $U_{\rm trans}$ are obtained by integrating the corresponding Poynting fluxes over time. Their normalized fractions characterize the redistribution of electromagnetic energy among the outgoing propagation directions. The complementary flux-balance quantity
\begin{equation}
\eta_{\rm dep}^{\rm est}
=
1-
\frac{U_{\rm back}+U_{\rm forward}+U_{\rm trans}}{U_{\rm inc}}
\label{eq:flux_balance_deposition}
\end{equation}
estimates the electromagnetic energy retained or deposited inside the integration boundary. It approximates the laser-energy absorption fraction when the residual electromagnetic-field energy inside that boundary and other unaccounted energy contributions are negligible. A fixed Fourier-band decomposition of $E_y$ separates forward- and backward-propagating optical components; their local backward-to-forward field-intensity ratio is used to compare optical-field buildup and is not interpreted as the global energy reflectivity.

Reflected-pulse arrival times are corrected for the geometry-dependent propagation time between the target surface and the left probe:
\begin{equation}
t_{\rm prop}=\frac{x_{\rm surf}-x_{\rm left}}{c},\qquad
\Delta\tau_{\rm refl}=\tau_{\rm peak,left}-t_{\rm prop}.
\end{equation}
Here $x_{\rm surf}$ is the longitudinal position of the detected target surface, $x_{\rm left}$ is the position of the left Poynting-flux probe, $c$ is the speed of light, and $\tau_{\rm peak,left}$ is the matched interaction time at which the primary backward-propagating pulse reaches its maximum at that probe. The corrected delay $\Delta\tau_{\rm refl}$ therefore removes the purely geometric propagation contribution before comparing different target designs. Independently, the early electron and total-particle energy fractions are calculated from stored scalar energies near $\tau\simeq\SI{50}{fs}$; these fractions are normalized to the instantaneous total particle-plus-field energy and are not absolute laser-absorption efficiencies. Secondary internal reflections are characterized separately so that prolonged reflected-field persistence is not assumed to control proton energy. Supplementary Sec. S11 provides the complete reflected-pulse timing, boundary-flux, and internal-reflection analyses.


\section{Optimization metrics}
\label{sec:metrics}

The proton cutoff energy $\Ecut$ is obtained from the final proton kinetic-energy spectrum as the largest energy bin satisfying
\begin{equation}
S(E)>10^{-6}S_{\max},
\label{eq:cutoff_definition}
\end{equation}
where $S_{\max}$ is the spectral maximum. This fixed relative threshold defines a reproducible spectral edge rather than the energy of the single most energetic macro-particle, which is more sensitive to sparse numerical tails and particle statistics. Lowering the threshold can only leave the cutoff unchanged or move it to a higher populated bin; any such change must therefore be interpreted together with the population carried by the terminal tail. Threshold sensitivity and cumulative high-energy-edge measures are quantified independently in the numerical validation and Supplementary Sec. S4.

The angular objective is reconstructed from the final joint $(p_x,p_y)$ proton histogram. For each bin,
\begin{equation}
E_k=\left[\sqrt{1+\frac{p_x^2+p_y^2}{m_s^2}}-1\right]m_s m_ec^2,
\label{eq:energy_reconstruction}
\end{equation}
with $m_s$ in electron-mass units and momenta in $m_ec$ units; this is therefore a reconstructed kinetic energy under the $p_z\simeq0$ assumption. Only forward-going bins with $p_x>0$ and reconstructed $E_k\geq\SI{5}{MeV}$ enter the primary beam-quality objective. Defining $\theta_i=\operatorname{atan2}(p_{y,i},p_{x,i})$ and histogram weights $w_i$, the angular centroid and weighted RMS divergence are
\begin{align}
\bar\theta&=\frac{\sum_iw_i\theta_i}{\sum_iw_i},\label{eq:pointing}\\
\DivH&=\left[\frac{\sum_iw_i(\theta_i-\bar\theta)^2}{\sum_iw_i}\right]^{1/2}.
\label{eq:divergence}
\end{align}
The optimization minimizes this centered RMS quantity, not the uncentered RMS angle.

The proton-only conversion efficiency is
\begin{equation}
\etaH=\frac{U_{\rm kin,H}}{E_{\rm in}},
\label{eq:etaH}
\end{equation}
where $U_{\rm kin,H}$ is the final kinetic energy of the complete hydrogen-ion population and $E_{\rm in}$ is the incident laser energy per unit out-of-plane length obtained from the integrated incoming Poynting flux at the common left probe. This is a proton-specific laser-to-proton conversion efficiency: carbon-ion and electron kinetic energies are not included in its numerator. It is therefore neither the total laser-absorption efficiency nor the total laser-to-ion conversion efficiency. Feasible designs satisfy
\begin{equation}
\etaH\geq0.09.
\label{eq:constraint}
\end{equation}
This campaign-level floor was fixed before Pareto selection to prevent apparent gains in cutoff or divergence from being retained when they were obtained at the expense of markedly reduced proton energy transfer. It is an operational feasibility constraint for the present design study, not a universal application or experimental threshold. All 82 optimization evaluations use the common production time $t=\SI{420}{fs}$; later-time, enlarged-domain, grid, and particle-number calculations are validation tests and do not redefine the frozen objectives.


\section{Constrained multi-objective Bayesian optimization}
\label{sec:optimization}

\begin{figure*}[!t]
\centering
\IfFileExists{figures/fig01_overview_final.pdf}{%
  \includegraphics[width=0.99\textwidth]{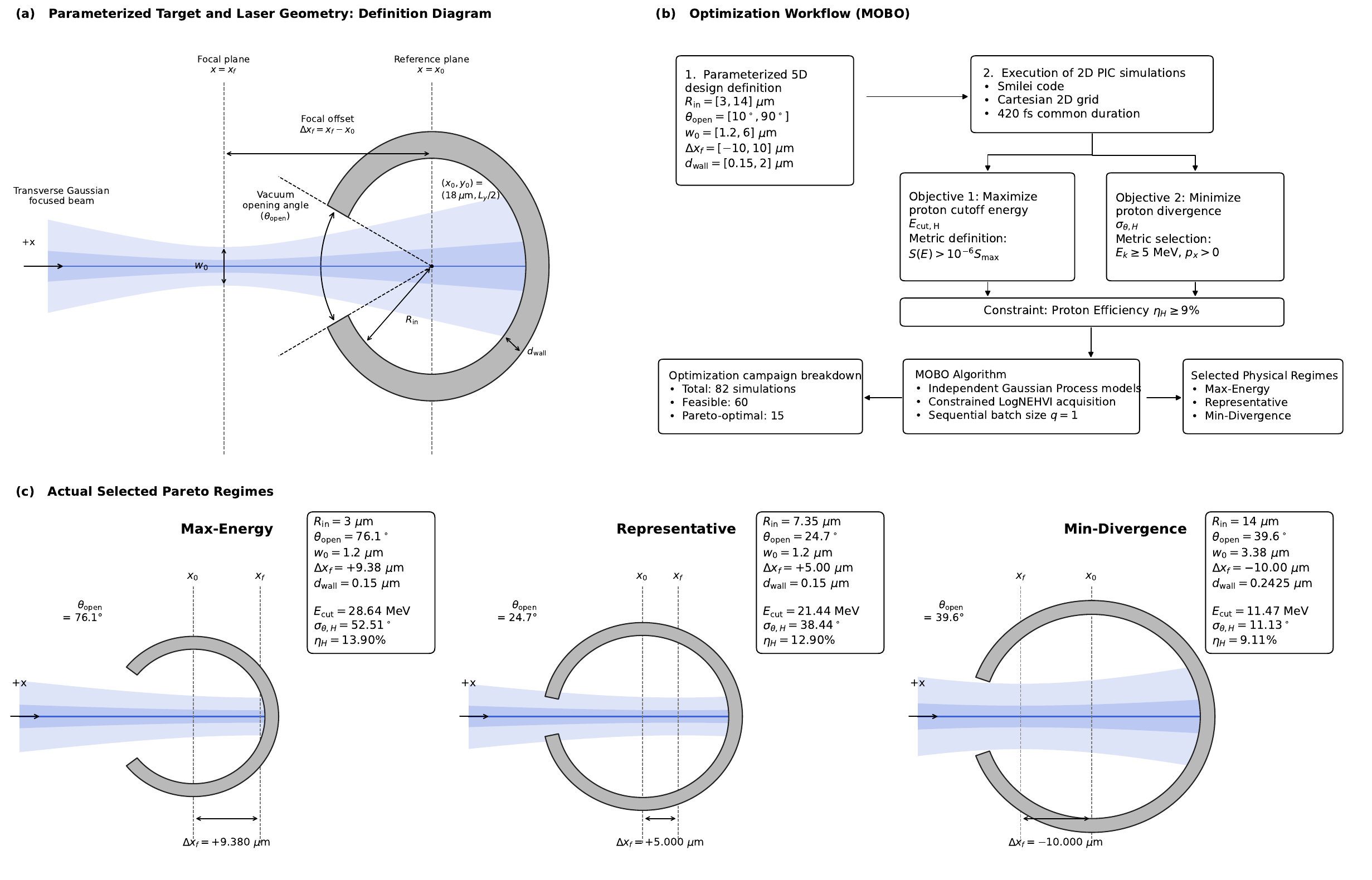}%
}{%
  \fbox{\parbox[c][7.0cm][c]{0.95\textwidth}{\centering
  Final Fig.~1 placeholder: copy \texttt{fig01\_overview\_final.pdf} to \texttt{figures/}.}}%
}
\caption{Parameterized C-shaped target, constrained-MOBO workflow, and selected Pareto regimes. (a) Definition of $R_{\rm in}$, $\theta_{\rm open}$, $d_{\rm wall}=R_{\rm out}-R_{\rm in}$, $w_0$, and the signed focal offset $x_f=x_0+\Delta x_f$. (b) Five-dimensional PIC--MOBO workflow maximizing proton cutoff energy and minimizing the centroid-referenced RMS divergence of forward protons with reconstructed $E_k\geq\SI{5}{MeV}$, subject to $\eta_H\geq9\%$. The final campaign contains 82 evaluations, 60 feasible designs, and 15 constrained Pareto-optimal solutions. (c) Max-Energy, Representative, and Min-Divergence designs used for the mechanism analysis; numerical parameter values and final performance are shown in the panel. Mini-schematics preserve qualitative geometric ordering and are not spatially scaled to the full simulation domain.}
\label{fig:target_workflow}
\end{figure*}

The design vector is
\[
\bm{x}=(R_{\rm in},\theta_{\rm open},w_0,\Delta x_f,d_{\rm wall}),
\]
with the geometric and laser conventions defined in Sec.~\ref{sec:numerical_model} and illustrated explicitly in Fig.~\ref{fig:target_workflow}(a). The physical objectives are to maximize the proton cutoff energy \(\Ecut\) and minimize the RMS divergence \(\DivH\) of forward protons with reconstructed \(E_k\geq\SI{5}{MeV}\), subject to \(\eta_H\geq0.09\). Internally, the objectives are represented in maximization form as \([\Ecut,-\DivH]\), while the constraint is written as \(0.09-\eta_H\leq0\).

Each design point is evaluated with one production-resolution PIC realization rather than a repeated-seed ensemble. Independent Gaussian-process surrogates are fitted to the two objectives and the feasibility constraint using normalized design coordinates and standardized outputs. For each output, the BoTorch \texttt{SingleTaskGP} model infers an effective homoskedastic observation-noise level through marginal-likelihood optimization; per-design noise variances are not supplied. Candidate selection is performed with a constrained logarithmic noisy expected-hypervolume-improvement strategy implemented in BoTorch~\cite{botorch2020,Daulton2021qNEHVI,Ament2023LogEI}. The noisy expected-hypervolume formulation provides a Bayesian treatment of uncertainty in the Pareto frontier, while the logarithmic reformulation improves numerical optimization of expected-improvement-type acquisitions in regimes where canonical acquisition values become very small. We use sequential candidate generation (\(q=1\)); acquisition optimization uses 12 multistart restarts and 512 raw samples. Feasibility labels refer consistently to the common production resolution and particle loading used across the campaign. Separate grid- and particle-number-refinement tests assess numerical sensitivity but are not repeated-seed estimates of stochastic PIC variability. The complete PIC--MOBO evaluation and constrained-selection sequence is summarized in Fig.~\ref{fig:target_workflow}(b).

The campaign progresses through an initial search domain $D_0$, a first expanded domain $D_1$, and a final expanded domain $D_2$, rather than assuming that the initial search box is sufficient. Starting from a 20-point design of experiments, 32 Bayesian-optimization evaluations in $D_0$ were followed by 20 extension evaluations in $D_1$ and 10 targeted boundary-confirmation evaluations in $D_2$, giving 82 production PIC evaluations in total. The final bounds of $D_2$ are
\[
R_{\rm in}\in[3,14]~\mu\mathrm m,\quad
\theta_{\rm open}\in[10,90]^\circ,\quad
w_0\in[1.2,6]~\mu\mathrm m,
\]
\[
\Delta x_f\in[-10,10]~\mu\mathrm m,\qquad
d_{\rm wall}\in[0.15,2]~\mu\mathrm m.
\]
The staged extensions were triggered by persistent non-dominated boundary occupancy, principally in \(R_{\rm in}\) and \(\Delta x_f\). The wall-thickness lower bound was deliberately not reduced below \SI{0.15}{\micro\meter}, because it is represented by only about six production-grid cells; instead, the two selected thin-wall regimes were subjected to a dedicated grid-resolution sensitivity analysis.

A temporary anti-boundary-collapse safeguard is applied only during acquisition optimization: repeated proposals on the same side of a variable move the corresponding acquisition bound inward by 5\% of the full physical range until the streak is broken. The physical design limits and previously evaluated data are never modified, and the safeguard is not treated as a physical prior. Optimization convergence is checked retrospectively with one fixed hypervolume reference applied to every historical campaign prefix; details of the stage-by-stage domain and hypervolume check are provided in Supplementary Secs.~S2--S3.

\section{Optimization results}
\label{sec:results_opt}

Of the 82 evaluated designs, 60 satisfy \(\eta_H\geq9\%\), and independent constrained non-dominated sorting identifies 15 feasible Pareto-optimal evaluated solutions. Across the complete adaptive dataset, cutoff energy and \(E_k\geq\SI{5}{MeV}\) RMS divergence are positively associated (Spearman \(\rho=0.696\)); the association strengthens within the feasible subset (\(\rho=0.787\)). These coefficients are descriptive of the adaptively sampled campaign rather than global sensitivity indices, but they directly confirm that energy and divergence behave as competing design objectives.

The staged domain extensions relieved the strongest earlier boundary pressure in \(R_{\rm in}\) and \(\Delta x_f\). The final front retains substantial accumulation at the lower \(d_{\rm wall}=\SI{0.15}{\micro\meter}\) bound, motivating the targeted thin-wall validation rather than extrapolation to unresolved thinner targets. The full domain-occupancy and fixed-reference hypervolume history is documented in Supplementary Secs.~S2--S3.

\begin{figure*}[!t]
\centering
\IfFileExists{figures/fig02_pareto_final.pdf}{%
  \includegraphics[width=0.98\textwidth]{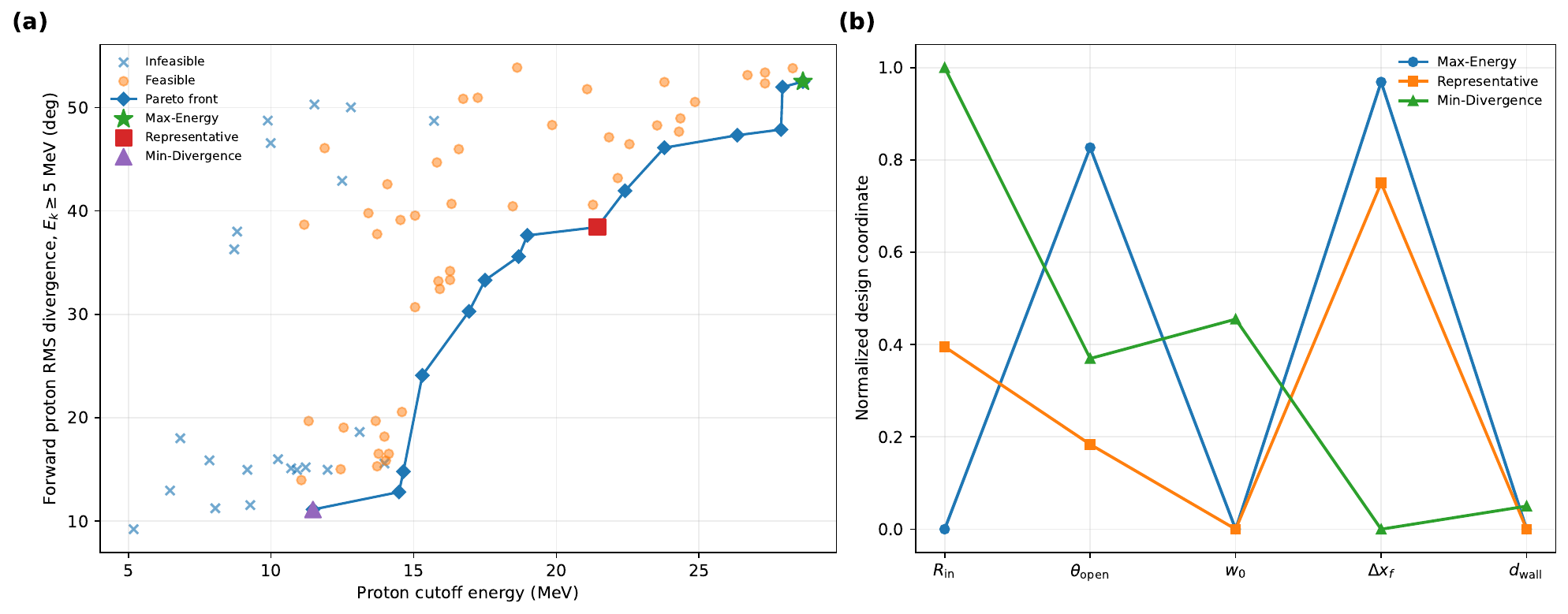}%
}{%
  \fbox{\parbox[c][5.0cm][c]{0.94\textwidth}{\centering
  Final Fig.~2 placeholder: copy \texttt{fig02\_pareto\_final.pdf} to \texttt{figures/}.}}%
}
\caption{Final constrained optimization landscape. (a) All 82 evaluations in cutoff-energy versus forward $E_k\geq\SI{5}{MeV}$ RMS-divergence space, with feasible/infeasible designs, the 15-point constrained Pareto front, and the three regimes selected for matched-stage analysis. (b) Normalized five-parameter fingerprints of the same three designs. Coordinate normalization is used only for comparing locations within the final search bounds and does not imply equal physical sensitivity.}
\label{fig:pareto_final}
\end{figure*}

The complete objective-space landscape is shown in Fig.~\ref{fig:pareto_final}(a), which separates feasible and infeasible evaluations, identifies the 15-point constrained Pareto front, and marks the three cases selected for detailed physical analysis. Figure~\ref{fig:pareto_final}(b) compares the normalized five-parameter coordinates of those selected designs, while Fig.~\ref{fig:target_workflow}(c) presents their physical target geometries and reported performance values. The Max-Energy design is
\begin{equation}
\begin{aligned}
&(R_{\rm in},\theta_{\rm open},w_0,\Delta x_f,d_{\rm wall})\\
&\qquad=(3.00,76.1^\circ,1.20,9.38,0.15),
\end{aligned}
\end{equation}
and reaches \(\Ecut=28.64~\mathrm{MeV}\), \(\DivH=52.51^\circ\), and \(\eta_H=13.90\%\). The Min-Divergence solution is
\begin{equation}
\begin{aligned}
&(R_{\rm in},\theta_{\rm open},w_0,\Delta x_f,d_{\rm wall})\\
&\qquad=(14.00,39.6^\circ,3.38,-10.00,0.24),
\end{aligned}
\end{equation}
and gives \(\Ecut=11.47~\mathrm{MeV}\), \(\DivH=11.13^\circ\), and \(\eta_H=9.11\%\). An equal-weight normalized ideal-distance rule applied only to the final Pareto set selects the Representative design,
\begin{equation}
\begin{aligned}
&(R_{\rm in},\theta_{\rm open},w_0,\Delta x_f,d_{\rm wall})\\
&\qquad=(7.35,24.7^\circ,1.20,5.00,0.15),
\end{aligned}
\end{equation}
with \(\Ecut=21.44~\mathrm{MeV}\), \(\DivH=38.44^\circ\), and \(\eta_H=12.90\%\). It is used as a reproducible compromise point rather than claimed as a unique knee or global optimum. Design coordinates quoted in the running text are rounded for readability; all simulations and Pareto calculations use the unrounded values stored in the campaign dataset.

\section{Matched-stage proton dynamics}
\label{sec:results_physics}

The three selected Pareto regimes were rerun in the enlarged $65\times50~\mu\mathrm{m}^2$ domain with diagnostics referenced to a case-specific interaction-time coordinate, \(\tau=t-t_{\rm ref}\). The reference \(t_{\rm ref}\) is the first 5\% crossing of the local incident laser-field envelope sampled \(0.5~\mu\mathrm m\) upstream of the detected target surface. It is therefore an operational low-amplitude interaction marker, not the pulse peak or literal infinitesimal first contact. The envelope peak follows the marker by approximately 23--24~fs in all three cases. These matched runs are used only for mechanism analysis; the optimization objectives remain frozen at the common production time \(t=\SI{420}{fs}\). Consequently, cutoff and divergence values extracted from the enlarged-domain matched runs are reported as mechanism-run cross-checks and do not replace the frozen campaign values used to construct the Pareto front.

\begin{figure*}[!t]
\centering
\IfFileExists{figures/fig03_proton_dynamics_final.pdf}{%
  \includegraphics[width=0.98\textwidth]{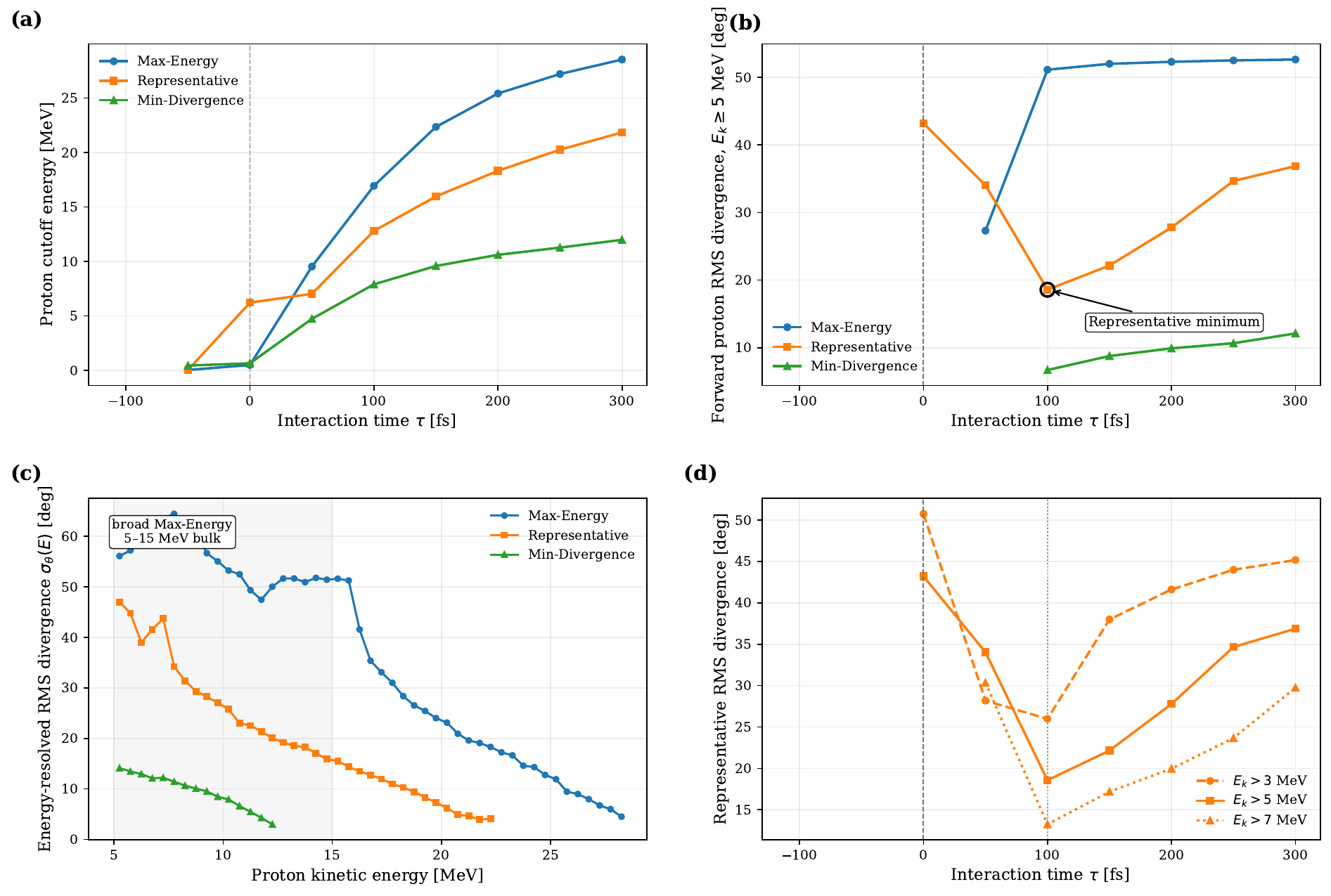}%
}{%
  \fbox{\parbox[c][6.0cm][c]{0.94\textwidth}{\centering
  Final Fig.~3 placeholder: copy \texttt{fig03\_proton\_dynamics\_final.pdf} to \texttt{figures/}.}}%
}
\caption{Matched-stage proton dynamics for the three selected Pareto regimes. (a) Proton cutoff energy and (b) forward $E_k\geq\SI{5}{MeV}$ RMS divergence versus the common interaction-time coordinate; the Representative design shows a transient collimation minimum near $\tau\simeq\SI{100}{fs}$. (c) Energy-resolved RMS angular width near $\tau\simeq\SI{300}{fs}$, showing a broad Max-Energy bulk but a progressively narrower terminal tail. (d) Representative threshold sensitivity for $E_k>3$, 5, and 7~MeV. Here $\tau=0$ is the frozen 5\% incident-envelope interaction marker, not the pulse peak.}
\label{fig:proton_dynamics}
\end{figure*}

Figure~\ref{fig:proton_dynamics}(a,b) shows that the three designs evolve into distinct proton regimes. At \(\tau\simeq300\) fs the Max-Energy case reaches a cutoff of 28.54~MeV with \(52.63^\circ\) forward \(E_k\geq\SI{5}{MeV}\) RMS divergence, whereas Min-Divergence reaches 11.98~MeV with only \(12.10^\circ\). The Representative case remains intermediate in final performance but has a non-monotonic angular history: its \(E_k\geq\SI{5}{MeV}\) divergence decreases from \(34.03^\circ\) at 50~fs to \(18.57^\circ\) at 100~fs before broadening to \(36.87^\circ\) at 300~fs while its cutoff continues to increase. As shown in Fig.~\ref{fig:proton_dynamics}(d), the same transient minimum survives the deliberately strict \(E_k>3\), 5, and 7~MeV sensitivity selections and fixed-energy comparisons over the reliable energetic range. It is therefore a genuine transient collimation of the energetic proton population rather than an artifact of the 5-MeV threshold.

The energy-resolved widths in Fig.~\ref{fig:proton_dynamics}(c) clarify why the large integrated divergence of the Max-Energy case does not imply that its entire energetic population is equally broad. Near \(\tau\simeq300\) fs, about 80.5\% of its \(E_k\geq\SI{5}{MeV}\) statistical weight lies between 5 and 15~MeV, where the angular widths remain large. By contrast, the terminal high-energy tail becomes progressively narrower: the pooled RMS divergence decreases from \(52.63^\circ\) for \(E_k\geq\SI{5}{MeV}\) to \(17.04^\circ\) for \(E_k>\SI{20}{MeV}\) and \(9.03^\circ\) for \(E_k>\SI{25}{MeV}\). Thus, the Max-Energy regime contains a broad, statistically dominant energetic bulk together with a substantially more collimated terminal tail. The detailed origin of this terminal-tail collimation is not resolved by the present mechanism diagnostics and is therefore kept separate from the global energy--divergence interpretation.

\begin{figure*}[!t]
\centering
\IfFileExists{figures/fig04_proton_spectra_phase_space.png}{%
  \includegraphics[width=0.99\textwidth]{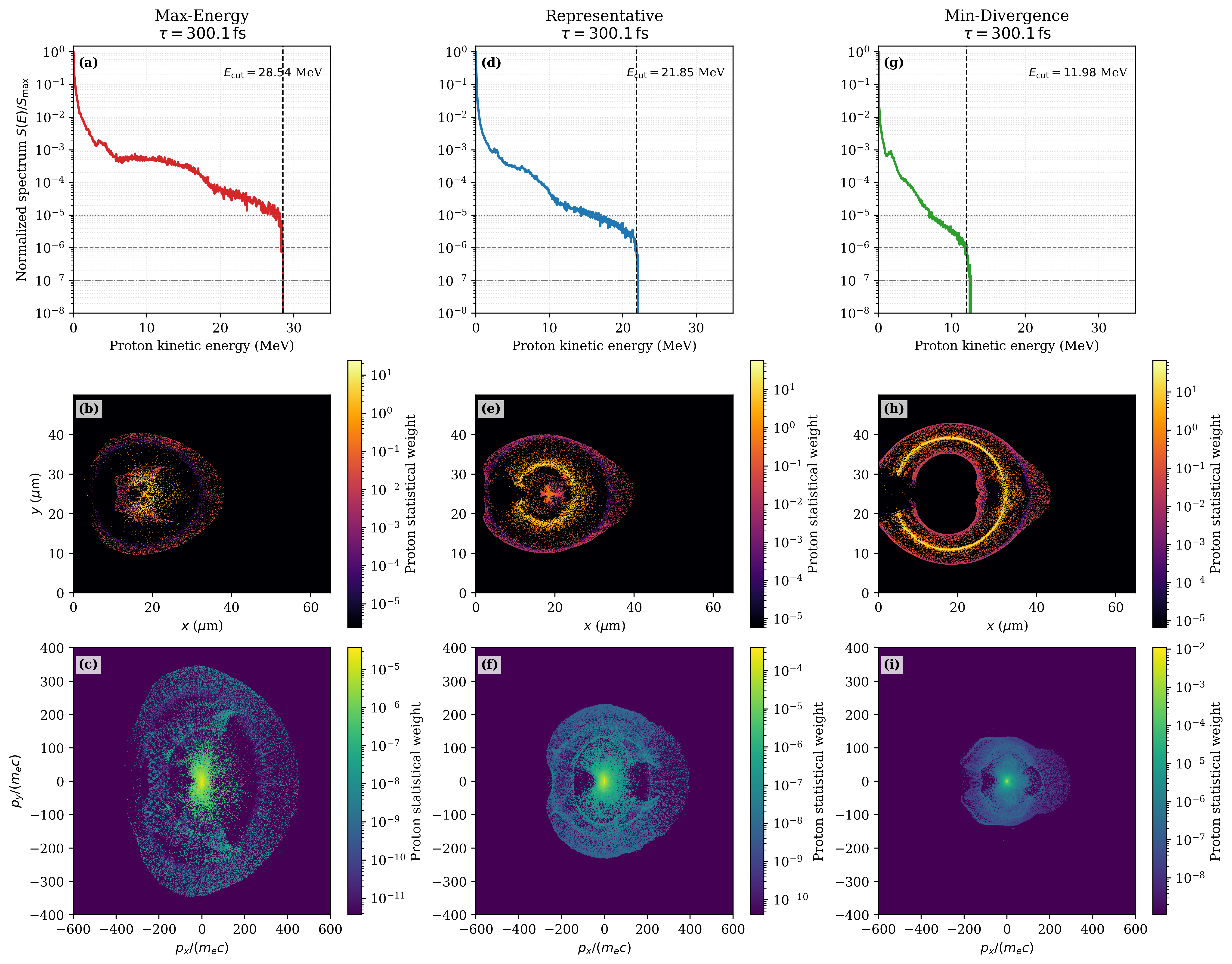}%
}{%
  \fbox{\parbox[c][7.0cm][c]{0.95\textwidth}{\centering
  Missing \texttt{figures/fig04\_proton\_spectra\_phase\_space.png}.}}%
}
\caption{Distribution-level proton comparison at the matched developed stage, $\tau\simeq\SI{300}{fs}$. Each column corresponds to the Max-Energy, Representative, or Min-Divergence design. Panels (a,d,g) show the normalized proton kinetic-energy spectra; horizontal lines mark relative spectral thresholds of $10^{-5}$, $10^{-6}$, and $10^{-7}$, and the vertical dashed lines indicate the primary $10^{-6}$ cutoff. Panels (b,e,h) show the instantaneous proton $x$--$y$ statistical-weight distributions, and panels (c,f,i) show the joint $p_x$--$p_y$ statistical-weight distributions. The spectra and momentum maps demonstrate that the optimized regimes differ at the distribution level rather than only through scalar objective values.}
\label{fig:proton_spectra_phase_space}
\end{figure*}

Figure~\ref{fig:proton_spectra_phase_space} complements the scalar and energy-resolved quantities in Fig.~\ref{fig:proton_dynamics} with direct distribution-level evidence. The stored kinetic-energy spectra verify the ordering of the terminal proton energies, while the spatial statistical-weight maps and joint $p_x$--$p_y$ distributions show how the three selected regimes differ in spatial expansion and transverse-momentum support at the matched late stage. These distributions confirm that the reported energy--divergence trade-off represents systematic changes in the energetic proton populations rather than isolated cutoff values or a few sparsely sampled high-energy particles.

The ion-energy ordering is not restricted to protons. At $\tau\simeq\SI{300}{fs}$, the threshold-defined total carbon-ion cutoff energies are 101.77, 51.59, and \SI{26.29}{MeV} for Max-Energy, Representative, and Min-Divergence, respectively. The Max-Energy carbon edge is stable over the tested $10^{-5}$--$10^{-7}$ relative thresholds, whereas the terminal tails of the other two cases are more threshold sensitive; the full carbon spectra and sensitivity values are reported in Supplementary Sec.~S13. Near the maximum early-sheath stage, the corresponding effective hot-electron slope temperatures follow the same ordering, 1.89, 1.63, and \SI{1.01}{MeV}, but are interpreted as nonthermal spectral-tail parameters rather than equilibrium temperatures.

Instantaneous proton spatial-energy moments provide an important negative result. The Representative momentum-space minimum at \(\tau\simeq100\) fs is not accompanied by contraction of the instantaneous spatial proton-energy distribution: its energy-weighted polar support continues to broaden from 50 to 200~fs. Consequently, the transient angular narrowing cannot be explained by a simple collapse of the propagated proton cloud and instead motivates analysis of the accelerating sheath relative to the dynamically evolving target surface. Detailed spatial-moment and threshold-sensitivity analysis are given in Supplementary Secs.~S4 and the matched-stage archive.

\section{Outer-sheath organization}
\label{sec:results_sheath}

\begin{figure*}[!t]
\centering
\IfFileExists{figures/fig05_outer_sheath_final.pdf}{%
  \includegraphics[width=0.98\textwidth]{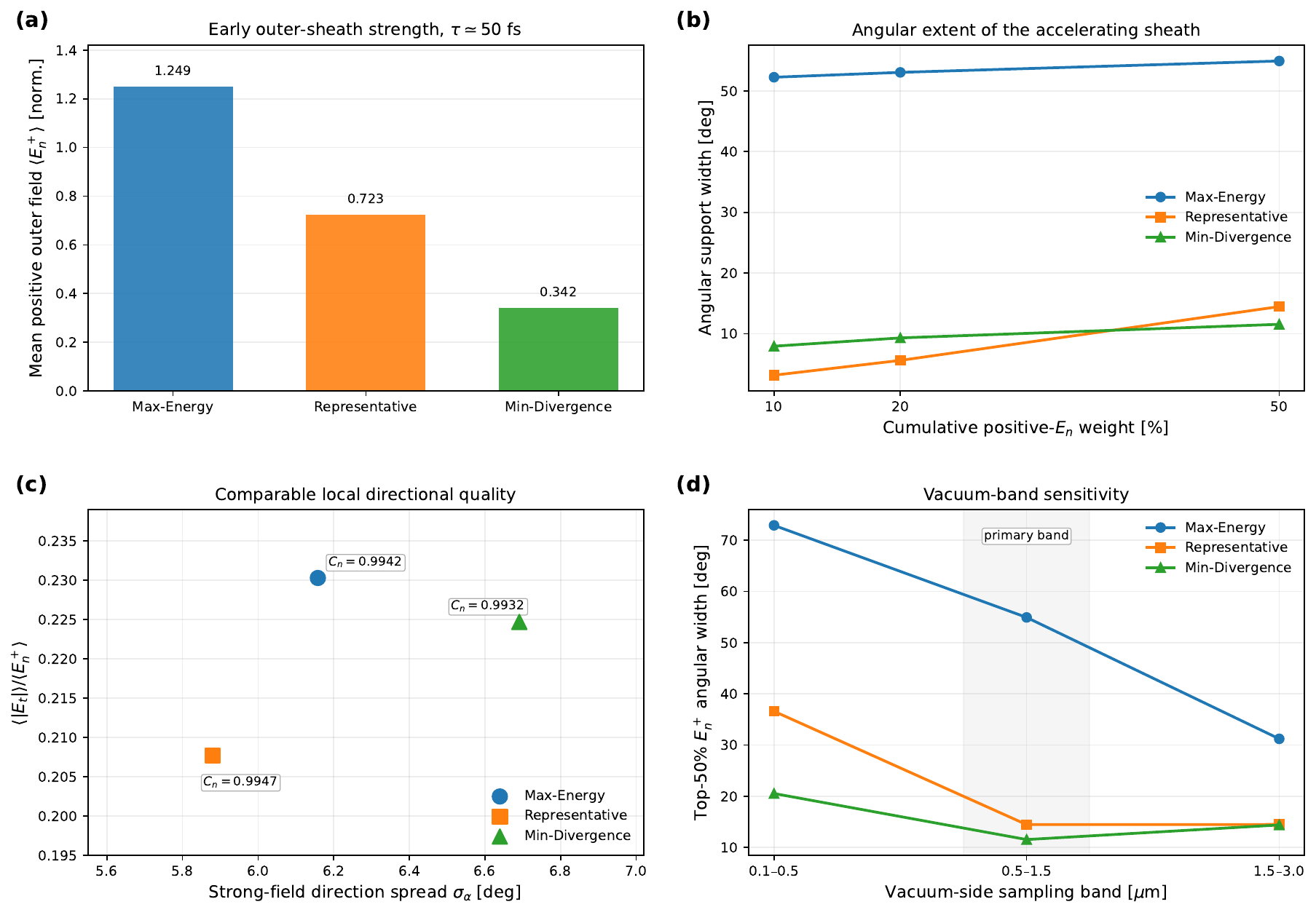}%
}{%
  \fbox{\parbox[c][6.2cm][c]{0.94\textwidth}{\centering
  Final Fig.~5 placeholder: copy \texttt{fig05\_outer\_sheath\_final.pdf} to \texttt{figures/}.}}%
}
\caption{Early outer-sheath organization at $\tau\simeq\SI{50}{fs}$. (a) Mean positive vacuum-facing normal field in the primary 0.5--\SI{1.5}{\micro\meter} band. (b) Angular support containing the top 10\%, 20\%, and 50\% of positive $E_n$ weight. (c) Local strong-field direction spread versus tangential-to-normal field ratio, with normal-alignment coherence $C_n$. (d) Top-50\% support-width sensitivity to the vacuum-side sampling band. The principal inter-case distinction is sheath strength and angular support, not local directional disorder; field-support width is not identified one-to-one with proton divergence.}
\label{fig:outer_sheath}
\end{figure*}

The dynamically reconstructed carbon interfaces allow the accelerating field to be resolved in a local vacuum-facing basis. The principal comparison uses the outer interface and the 0.5--\SI{1.5}{\micro\meter} vacuum-side band near the early strong-sheath stage \(\tau\simeq\SI{50}{fs}\). As shown in Fig.~\ref{fig:outer_sheath}(a), the mean positive normal field is 1.249, 0.723, and 0.342 in normalized units for Max-Energy, Representative, and Min-Divergence, respectively. Their local directional quality, quantified in Fig.~\ref{fig:outer_sheath}(c), is nevertheless similar: the strong-field direction spreads are \(6.16^\circ\), \(5.88^\circ\), and \(6.69^\circ\), and the normal-alignment coherences are all about 0.99. The Max-Energy regime is therefore not distinguished primarily by a locally less-normal sheath.

The dominant inter-case difference is the angular support over which a strong outward sheath is sustained. Figure~\ref{fig:outer_sheath}(b) shows that the top-50\% positive-\(E_n\) support widths are \(54.93^\circ\), \(14.47^\circ\), and \(11.55^\circ\) for Max-Energy, Representative, and Min-Divergence. The sampling-band comparison in Fig.~\ref{fig:outer_sheath}(d) shows that the absolute width depends on vacuum sampling distance, but the same broad-to-localized ordering is retained in all three tested bands. This field ordering precedes and is consistent with the subsequently developed energetic-proton divergence: near \(\tau\simeq100\) fs the corresponding \(E_k\geq\SI{5}{MeV}\) proton widths are approximately \(51.13^\circ\), \(18.57^\circ\), and \(6.69^\circ\). The field-support and proton-divergence measures are not equated pointwise; the former is a surface-polar diagnostic at one time, whereas the latter integrates particle histories.

The Representative case provides a temporal consistency check. Its early coherent sheath degrades as the interaction develops, while the energetic-proton divergence reaches its minimum later. A discrete time-delay analysis is consistent with a finite response/accumulation interval of order 50--100~fs, but the small number of temporally correlated snapshots does not support a uniquely measured 100-fs response time.

Two safeguards delimit the field-level interpretation. First, the broad Max-Energy sheath does not identify a narrow strongest-\(E_n\) sector that could explain its collimated terminal proton tail; that tail therefore remains unresolved. Second, the Max-Energy interface tracker undergoes a late branch switch near \(\tau\simeq300\) fs, so those late interface-based quantities are excluded from causal interpretation. The geometry calibration, distance-band sensitivity, and tracker-reliability analysis are documented in Supplementary Sec.~S8.

\section{Optical coupling, electron redistribution, and charge separation}
\label{sec:results_upstream}

\begin{figure*}[!t]
\centering
\IfFileExists{figures/fig06_upstream_mechanism_final.pdf}{%
  \includegraphics[width=0.98\textwidth]{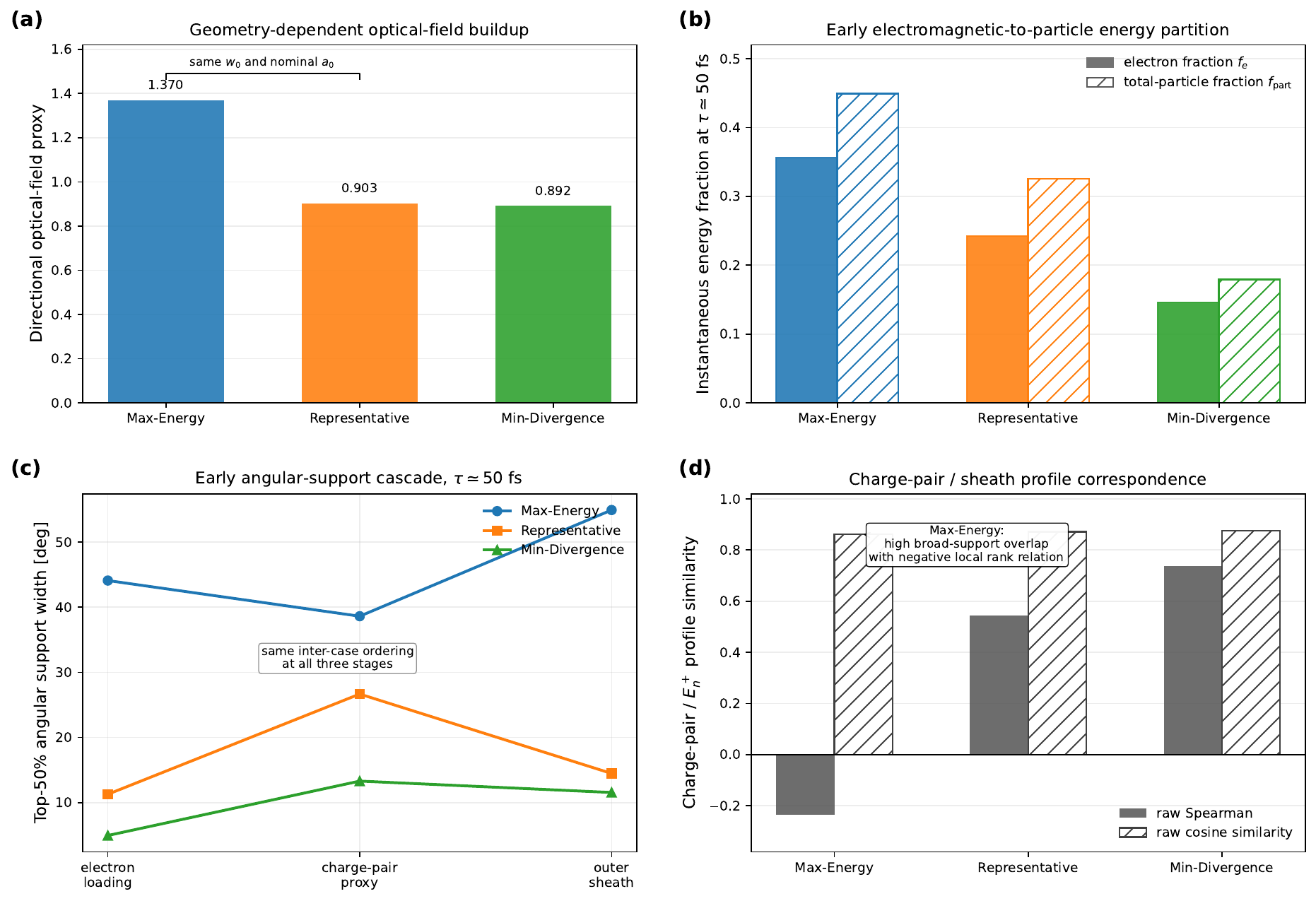}%
}{%
  \fbox{\parbox[c][6.2cm][c]{0.94\textwidth}{\centering
  Final Fig.~6 placeholder: copy \texttt{fig06\_upstream\_mechanism\_final.pdf} to \texttt{figures/}.}}%
}
\caption{Early pathway from optimized laser--target coupling to outer-sheath organization near $\tau\simeq\SI{50}{fs}$. (a) Local backward-to-forward optical-field intensity ratio. Max-Energy and Representative have the same $w_0$ and nominal $a_0$, so their different ratios isolate the effect of geometry and focal position; this field ratio is not an energy reflectivity. (b) Instantaneous electron and total-particle fractions of the internally stored energy, showing that the optical ordering is accompanied by the same Max-Energy $>$ Representative $>$ Min-Divergence particle-energization ordering. (c) Top-50\% angular-support widths of electron kinetic-energy loading, the charge-pair proxy $S_{\rm pair}$, and the positive outer-sheath field $E_n^+$. Their common broad-to-localized ordering connects broad, strong early organization in Max-Energy to its high-energy/broad-divergence proton regime and localized organization in Min-Divergence to its lower-divergence/lower-energy regime. (d) Raw $S_{\rm pair}$--$E_n^+$ Spearman correlation and cosine similarity. Max-Energy retains high broad-profile similarity but negative local rank correspondence, demonstrating that the charge--field connection is non-local and regime scale rather than pointwise.}
\label{fig:upstream_mechanism}
\end{figure*}

The proton and sheath results establish \emph{what} differs among the three optimized regimes, but not \emph{why} those differences arise. Figure~\ref{fig:upstream_mechanism} addresses this question by following four consecutive steps of the early interaction near \(\tau\simeq\SI{50}{fs}\): optical-field buildup, transfer of energy to particles, angular organization of electrons and charge separation, and formation of the outer accelerating sheath. The purpose of the figure is not merely to show four correlated observables; it is to test a physically ordered pathway from the optimized target--laser configuration to the later proton energy--divergence regime.

\paragraph{Geometry-dependent optical-field buildup---Fig.~\ref{fig:upstream_mechanism}(a).}
Panel (a) compares the fixed-band local backward-to-forward optical-field intensity ratio \(E_{y,-x}^{2}/E_{y,+x}^{2}\). For the same three selected campaign evaluations used to define the Pareto objectives and proton efficiencies, integration of \(PoyX\) through the common left control surface gives incident energies of 1504.83, 1504.79, and \(1507.16~\mathrm{J\,cm^{-1}}\) for Max-Energy, Representative, and Min-Divergence, respectively. Their full range, evaluated as \(100(E_{\max}-E_{\min})/\overline{E}_{\rm in}\), is only 0.158\%, ruling out unequal delivered laser energy as the origin of the observed inter-case differences. More importantly, Max-Energy and Representative have the same waist, \(w_0=\SI{1.2}{\micro\meter}\), and the same nominal \(a_0\simeq15.81\), but their optical-field buildup ratios are 1.370 and 0.903, respectively. This same-intensity comparison isolates the effect of the optimized cavity geometry and focal position: Max-Energy produces substantially stronger local counter-propagating/multi-wave field buildup. The plotted ratio is a local field diagnostic, not an energy reflectivity, so a value above unity does not imply more than 100\% reflected energy.

\paragraph{Optical-field buildup is accompanied by enhanced early particle energization---Fig.~\ref{fig:upstream_mechanism}(b).}
Panel (b) shows the instantaneous fractions of the internally stored energy carried by electrons and by all particles. The electron-energy fractions decrease from 0.356 to 0.243 and 0.146, and the total-particle fractions decrease from 0.449 to 0.326 and 0.179, for Max-Energy, Representative, and Min-Divergence, respectively. Thus, the optical ordering in panel (a) is accompanied by the same ordering in early particle energization. These quantities are instantaneous energy fractions inside the simulation domain, not absolute laser-absorption efficiencies; their role is to establish a consistent inter-case energy-transfer ordering.

\paragraph{Angular organization of the energized population and outer sheath---Fig.~\ref{fig:upstream_mechanism}(c).}
Panel (c) compares the top-50\% angular-support widths of three successive responses: electron kinetic-energy loading, the charge-pair proxy \(S_{\rm pair}\), and the positive outer-sheath field \(E_n^+\). For Max-Energy these structures extend over broad angular sectors, with widths of \(44.10^\circ\), \(38.62^\circ\), and \(54.93^\circ\), respectively. Representative is more localized, with corresponding widths of \(11.28^\circ\), \(26.67^\circ\), and \(14.47^\circ\), while Min-Divergence is the most localized overall, with \(4.94^\circ\), \(13.30^\circ\), and \(11.55^\circ\). The three widths are not expected to coincide numerically because they describe different physical quantities. The important result is their common broad-to-localized regime ordering. This provides the central connection to the proton trade-off: Max-Energy combines stronger energization with a sheath acting over a wide part of the curved surface, favoring higher proton energy but a broad energetic-proton population; Min-Divergence confines the strong accelerating structure to a much narrower sector, favoring lower divergence at the cost of spectral extent. Representative lies between these limits.

\paragraph{The charge-separation--sheath relation is non-local---Fig.~\ref{fig:upstream_mechanism}(d).}
Panel (d) is a safeguard against over-interpreting panel (c). It compares the angular profiles of \(S_{\rm pair}\) and \(E_n^+\) using raw Spearman rank correlation and cosine similarity. Representative and Min-Divergence have positive raw Spearman coefficients of 0.544 and 0.737, whereas Max-Energy has a negative coefficient of -0.235 despite a high broad-profile cosine similarity of 0.861. Therefore, similar regime-scale breadth does not imply that the largest charge-pair value and the largest sheath field occur at the same angular location. This is physically reasonable because the electrostatic field is generated non-locally by the extended charge distribution. The supported statement is consequently that charge-separation \emph{strength and angular organization} are consistent with the sheath regime, not that \(S_{\rm pair}(\phi)\propto E_n(\phi)\) point by point. A fixed \(3^\circ\) smoothing sensitivity test reported in Supplementary Sec.~S10 leads to the same conclusion.

Taken together, Fig.~\ref{fig:upstream_mechanism}(a--d) supports the following regime-scale mechanism:
\[
\begin{aligned}
&\text{optimized geometry and focus}
\rightarrow \text{local optical-field buildup}
\rightarrow \text{early particle energization}\\
&\qquad\rightarrow \text{angular organization of electrons and charge separation}\\
&\qquad\rightarrow \text{outer-sheath strength and support}
\rightarrow \text{proton energy--divergence regime}.
\end{aligned}
\]
The evidence supports the ordering of these stages across the three optimized designs, but it does not constitute particle-by-particle causal tracking or a local constitutive law. A separate electron-loading/sheath profile test in Supplementary Sec.~S9 reaches the same non-local conclusion: Representative and Min-Divergence show positive local rank correspondence, while Max-Energy combines broad-profile overlap with a negative raw rank coefficient.

Finally, two timing diagnostics rule out a simpler interpretation based only on optical trapping duration. After subtracting the geometry-dependent wall-to-probe propagation time, the corrected primary-reflection delays are 24.142, 23.221, and 23.639~fs, spanning only 0.921~fs. Moreover, Representative, rather than Max-Energy, has the largest secondary-to-primary reflected-envelope ratio. Internal reflections therefore contribute to geometry-dependent field buildup, but neither the corrected return time nor the persistence of secondary reflections controls the proton-energy ordering by itself. Supplementary Secs.~S9--S11 provide the complete electron-locality, charge-reconstruction, timing, and energy-escape analyses.

\paragraph{Cross-diagnostic regime fingerprint---Fig.~\ref{fig:mechanism_robustness}(a).}
Figure~\ref{fig:mechanism_robustness} first condenses the preceding diagnostics into one four-panel synthesis before testing their numerical robustness. Panel (a) compares the three selected regimes across the stored optical, particle-energy, charge, sheath, and final proton observables. Each row is normalized independently by the largest value of that observable among the three cases, so the panel preserves inter-case ordering but not absolute units or a universal larger-is-better meaning. In particular, the Max-Energy divergence entry equals unity because Max-Energy has the largest raw \(E_k\geq\SI{5}{MeV}\) RMS divergence, \(52.51^\circ\), compared with \(38.44^\circ\) for Representative and \(11.13^\circ\) for Min-Divergence. This value is neither a divergence of \(1^\circ\) nor a beam-quality score. Read in this way, the fingerprint makes the trade-off explicit: Max-Energy combines the largest cutoff and strongest/broadest upstream response with the broadest energetic-proton population; Min-Divergence shows the most localized upstream organization and the smallest final divergence; Representative lies between these endpoints. Panel (a) is therefore a compact synopsis of independently reported measurements, not a new fit or sensitivity analysis.

\paragraph{Physical pathway assembled from the fingerprint---Fig.~\ref{fig:mechanism_robustness}(b).}
Panel (b) then reads the contrasts summarized in panel (a) from upstream cause to downstream response. The matched-\(w_0\), matched-\(a_0\) Max-Energy/Representative comparison controls the nominal waist and intensity while the target geometry and focal offset remain different; it therefore tests their combined influence on optical buildup rather than isolating geometry alone. The corresponding optical ratios are 1.370 and 0.903. The subsequent electron-energy fractions follow the three-case sequence \(0.356>0.243>0.146\); the positive-sheath top-50\% supports likewise contract from \(54.93^\circ\) to \(14.47^\circ\) and \(11.55^\circ\). These upstream differences culminate in proton cutoffs of 28.64, 21.44, and 11.47~MeV and \(E_k\geq\SI{5}{MeV}\) RMS divergences of \(52.51^\circ\), \(38.44^\circ\), and \(11.13^\circ\), respectively. The arrows in panel (b) connect these regime-level orderings; they do not assert pointwise causality. Its safeguard boxes retain the two limits established above: longer optical recirculation is not a monotonic proton-energy control variable, and charge separation does not map pointwise onto the sheath field.

\section{Numerical validation}
\label{sec:validation}

The optimization compares all 82 designs at the common production time \(t=\SI{420}{fs}\); later-time runs are used only as robustness checks and do not redefine the objectives. Targeted validation was performed on the Max-Energy endpoint and the Representative compromise, with full numerical tables and threshold-sensitivity analysis in Supplementary Secs.~S4--S7 and S12.

For Max-Energy, the production \(\lambda_0/32\) grid gives \((\Ecut,\DivH,\eta_H)=(28.64~\mathrm{MeV},52.5^\circ,13.90\%)\). Refinements near \(\lambda_0/40\) and \(\lambda_0/50\) give \((29.56~\mathrm{MeV},52.6^\circ,13.94\%)\) and \((29.41~\mathrm{MeV},52.5^\circ,14.20\%)\), respectively. Between the two finest grids the cutoff changes by only -0.52\% and the divergence by -0.08\%; cumulative high-energy-edge metrics likewise remain within about 2\%. An independent Representative refinement from \(\lambda_0/32\) to \(\sim\lambda_0/40\) changes the cutoff by -0.71\%, divergence by +1.55\%, and proton efficiency by -1.72\%.

Transverse-domain enlargement changes the Max-Energy cutoff, divergence, and efficiency at 420~fs by +1.25\%, +0.35\%, and +0.30\%, respectively. Extending the larger-domain run to about 600~fs shows that the finite-box cutoff continues to evolve and reaches a stored maximum near 560~fs before decreasing, whereas the high-energy RMS divergence remains essentially unchanged. The reported cutoff is therefore a fixed-time optimization observable, not an asymptotic proton energy.

Particle-number refinement from 9 to 16 and 25 particles per cell leaves the divergence stable and brings the cutoff toward a particle-number plateau: the 16-to-25 PPC cutoff change is only 0.51\%. Integrated proton efficiency and the energetic-proton fraction retain a larger residual PPC dependence and are not described as fully particle-number converged.

The two selected \(d_{\rm wall}=\SI{0.15}{\micro\meter}\) regimes were additionally repeated at \(\lambda_0/40\) for the shell-resolved mechanism. Near the pre-registered strong-sheath stage \(\tau\simeq\SI{50}{fs}\), the positive-sheath strength changes by -1.62\% for Max-Energy and +2.52\% for Representative; the corresponding top-50\% sheath angular widths change by only \(-0.20^\circ\) and \(-0.03^\circ\). Charge-pair amplitudes show larger but bounded changes of +12.37\% and +17.58\%, and their angular widths change by \(+0.94^\circ\) and \(+0.35^\circ\). All pre-registered quantitative gates pass and the Max-Energy $>$ Representative ordering is retained. Detector-threshold recalibration changes the refined pair amplitudes by less than 0.8\%.

\begin{figure*}[!t]
\centering
\IfFileExists{figures/fig07_mechanism_robustness_final.pdf}{%
  \includegraphics[width=0.99\textwidth]{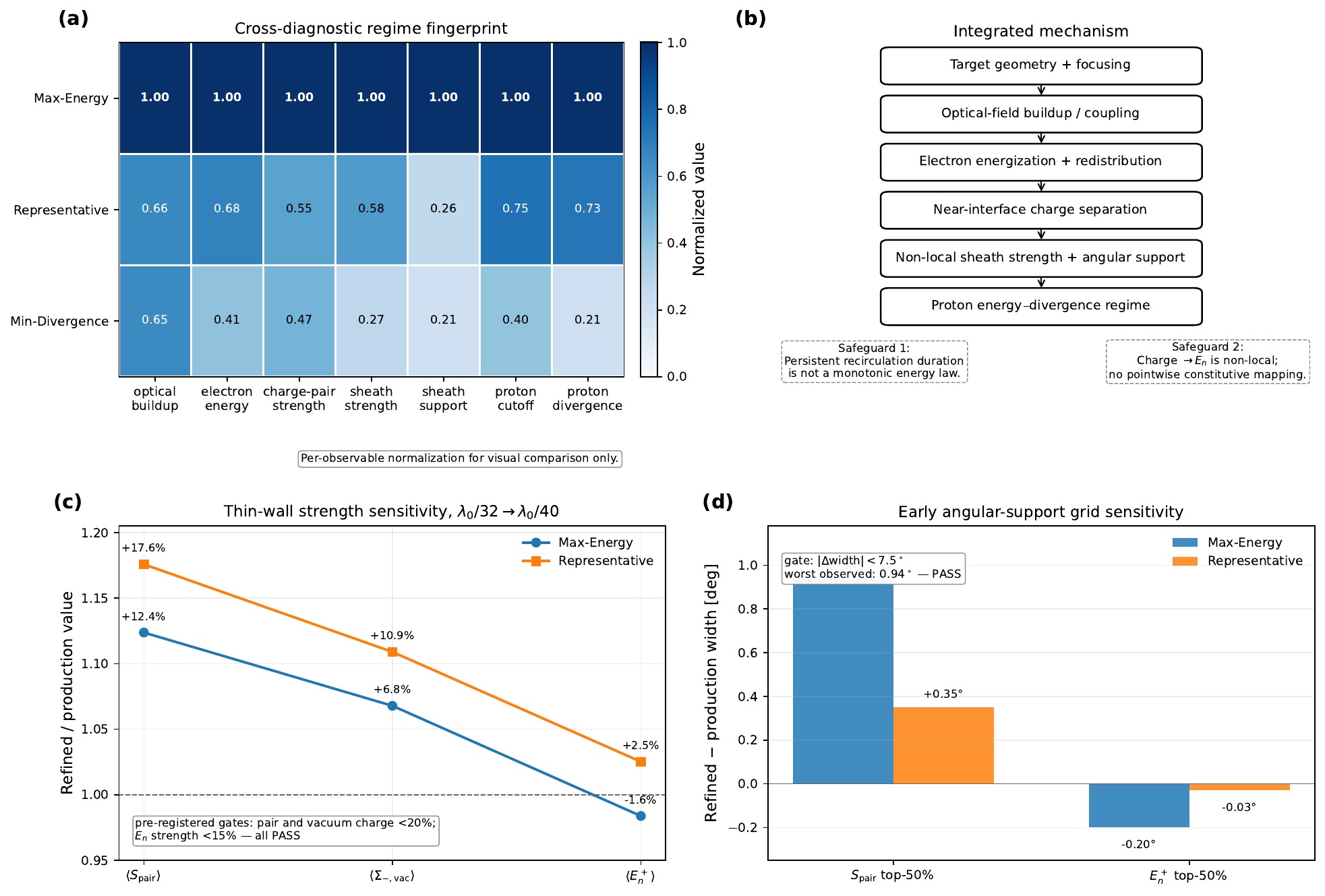}%
}{%
  \fbox{\parbox[c][6.5cm][c]{0.95\textwidth}{\centering
  Final Fig.~7 placeholder: copy \texttt{fig07\_mechanism\_robustness\_final.pdf} to \texttt{figures/}.}}%
}
\caption{Integrated regime comparison, physical interpretation, and targeted numerical robustness. (a) Cross-diagnostic fingerprint of the three selected regimes. Each observable is normalized independently to its largest value among the three cases; unity therefore denotes the largest raw value in that row, not necessarily the most desirable outcome. In particular, the Max-Energy divergence entry is unity because that case has the largest raw RMS divergence. The panel is a compact synopsis of previously quantified observables, not a correlation fit or sensitivity metric. (b) The same contrasts are read from upstream to downstream as a regime-scale physical pathway: target geometry and focusing modify optical-field buildup and coupling, electron energization and redistribution, near-interface charge-separation organization, non-local outer-sheath strength and angular support, and finally the proton energy--divergence regime. The safeguard boxes exclude a monotonic recirculation-duration law and a pointwise charge--field mapping. (c) Thin-wall $\lambda_0/32\rightarrow\lambda_0/40$ refined-to-production amplitude ratios near $\tau\simeq\SI{50}{fs}$ test the strength component of this interpretation. (d) Corresponding changes in the top-50\% angular-support widths test its broad-versus-localized component. All pre-registered gates pass; panels (c,d) establish targeted resolution robustness rather than Richardson-style asymptotic convergence.}
\label{fig:mechanism_robustness}
\end{figure*}

\paragraph{Amplitude robustness---Fig.~\ref{fig:mechanism_robustness}(c).}
After the regime fingerprint and mechanism summary in Fig.~\ref{fig:mechanism_robustness}(a,b), panel (c) tests whether the shell-resolved amplitudes supporting that interpretation survive the targeted thin-wall refinement. It reports the \(\lambda_0/40\)-to-\(\lambda_0/32\) ratios for the charge-pair, negative-vacuum-charge, and positive outer-sheath measures at \(\tau\simeq\SI{50}{fs}\). The charge-related changes remain below the pre-registered 20\% limits, the positive-sheath change is at most 2.52\%, and the Max-Energy \(>\) Representative strength ordering is retained.

\paragraph{Angular-support robustness---Fig.~\ref{fig:mechanism_robustness}(d).}
Panel (d) separately tests the angular quantities used to distinguish broad from localized acceleration. The largest absolute refinement-induced change is \(0.94^\circ\), well below the \(7.5^\circ\) gate, and the broad Max-Energy versus localized Representative ordering remains unchanged. Panels (c) and (d) therefore validate, respectively, the amplitude and angular-support components of the mechanism summarized in panels (a) and (b).

This last test is a targeted grid-sensitivity validation, not a Richardson/asymptotic convergence study. The early regime-scale charge/sheath mechanism is therefore treated as robust, while exact shell-integrated charge amplitudes remain \emph{resolution-checked} rather than formally converged. Late weak-sheath Max-Energy angular widths are more resolution sensitive and are outside this robustness claim.

\section{Discussion}
\label{sec:discussion}

The optimization and matched-stage diagnostics support a design-resolved interpretation of the energy--divergence trade-off. The 15-point constrained Pareto set demonstrates that high proton energy, low divergence, and useful conversion efficiency do not collapse to a single operating point within the investigated domain. The Max-Energy and Min-Divergence endpoints are not merely different scalar outcomes of the same dynamics; they develop distinct optical, electron, charge, sheath, and proton organizations. The Representative case adds a transient regime in which energetic protons become substantially more collimated around \(\tau\simeq100\) fs before broadening later.

The feasibility constraint uses the proton-specific efficiency \(\eta_H=U_{\rm kin,H}/E_{\rm in}\), not the total ion efficiency. To maintain a single quantitative basis, all efficiencies reported here are taken from the same three selected campaign evaluations used for the Pareto objectives. In each case, \(E_{\rm in}\) is obtained by integrating the incident \(PoyX\) signal through the common left control surface, and the species kinetic energies are evaluated at the common production time. For Max-Energy, Representative, and Min-Divergence, respectively, the laser-to-proton efficiencies are 13.90\%, 12.90\%, and 9.11\%. The corresponding laser-to-carbon efficiencies are 31.45\%, 22.37\%, and 13.42\%, giving total laser-to-ion efficiencies
\[
\eta_{\rm ions}=\frac{U_{\rm kin,H}+U_{\rm kin,C}}{E_{\rm in}}
\]
of 45.35\%, 35.28\%, and 22.53\%. These totals are evaluated from the unrounded species energies and incident energies; consequently, the separately displayed two-decimal proton and carbon percentages need not sum exactly in their last reported digit. Thus, the Max-Energy regime enhances energy transfer to both ion species rather than only extending the proton spectral edge. These conversion efficiencies are final-time kinetic-energy fractions of the indicated ion populations; they are not total laser-absorption efficiencies, because energy retained in electrons and electromagnetic fields or lost through the simulation boundaries is not included in their numerators. The total-ion values are reported here as a complementary physical measure and do not retrospectively alter the proton-specific feasibility constraint used in the optimization.

The RMS divergence calculated for all forward protons with \(E_k\geq\SI{5}{MeV}\) characterizes the complete selected energetic population and should not be interpreted as the divergence of protons near the spectral cutoff. In the Max-Energy case, approximately 80.5\% of the selected statistical weight lies between 5 and 15~MeV, where the angular distribution is broad; this population therefore dominates the \(E_k\geq\SI{5}{MeV}\) RMS divergence. In contrast, progressively higher-energy selections are substantially narrower: the pooled RMS divergence decreases from \(52.63^\circ\) for \(E_k\geq\SI{5}{MeV}\) to \(17.04^\circ\) for \(E_k>\SI{20}{MeV}\) and \(9.03^\circ\) for \(E_k>\SI{25}{MeV}\). The Max-Energy regime therefore consists of a broad, statistically dominant energetic bulk together with a less-populated but more collimated terminal tail. The present diagnostics do not identify the detailed acceleration history responsible for this terminal-tail collimation, so it is not included in the proposed mechanism for the global energy--divergence trade-off.

The Representative case demonstrates that momentum-space collimation does not require contraction of the instantaneous spatial proton distribution. Although its forward \(E_k\geq\SI{5}{MeV}\) RMS divergence decreases from \(34.03^\circ\) at \(\tau\simeq\SI{50}{fs}\) to \(18.57^\circ\) at \(\tau\simeq\SI{100}{fs}\), the spatial angular width of its proton kinetic-energy content increases over the same interval. The transient angular minimum must therefore be associated with the organization of the accelerating field rather than with simple spatial compression of the proton cloud. Surface-resolved analysis shows that the strong early sheath remains similarly aligned with the local target normal in all three regimes, but differs substantially in strength and angular support. Max-Energy sustains the strongest sheath over the broadest surface sector, whereas Representative and Min-Divergence restrict strong acceleration to progressively narrower sectors. Because different portions of the curved surface have different local normals, this broad-to-localized sheath ordering provides a field-level explanation for the subsequently developed broad-to-narrow energetic-proton angular regimes. The comparison establishes a regime-scale correspondence and does not equate the instantaneous sheath width with the later proton divergence.

The electron and charge diagnostics identify the processes preceding outer-sheath formation. Across the three designs, the angular support of electron kinetic-energy loading, the charge-separation proxy \(S_{\rm pair}\), and the positive outer-sheath field \(E_n^+\) follows the same broad-to-localized ordering from Max-Energy to Representative and Min-Divergence. This agreement supports a regime-scale connection between electron redistribution, charge separation, and sheath organization. It does not, however, imply a pointwise proportionality between \(S_{\rm pair}(\phi)\) and \(E_n(\phi)\). In Max-Energy, the two profiles have a high cosine similarity because both extend over a broad and substantially overlapping surface sector, but their negative Spearman coefficient shows that their local maxima and angular rank ordering do not coincide. This absence of pointwise correspondence persists after fixed-angle smoothing, as expected for an electrostatic field generated non-locally by an extended charge distribution. The supported conclusion is therefore that the optimized designs control the overall strength and angular extent of charge separation, which in turn organizes the regime-scale structure of the outer sheath.

The optical-field analysis moves the mechanism one step further upstream. Although the incident laser energies are nearly identical, the optimized geometry and focal position strongly modify the local superposition of forward- and backward-propagating fields and the subsequent early particle-energy partition. The comparison between Max-Energy and Representative is particularly informative because both have the same \(w_0\) and nominal \(a_0\), yet Max-Energy exhibits substantially stronger local counter-propagating field buildup. This result supports an important role for geometry-dependent internal reflection, optical recirculation, and transient field confinement in enhancing local coupling. However, neither the corrected primary-reflection delay nor the persistence of the secondary reflected envelope follows the proton-cutoff ordering. Internal recirculation therefore contributes to the coupling pathway, but its duration or persistence alone is not the scalar quantity controlling the final proton energy.

The strongest mechanism supported by the present dataset is consequently
\begin{equation}
	\begin{aligned}
		&\text{target geometry + focusing} \\
		&\quad\rightarrow \text{optical-field buildup/coupling} \\
		&\quad\rightarrow \text{electron energization/redistribution} \\
		&\quad\rightarrow \text{near-interface charge-separation organization} \\
		&\quad\rightarrow \text{non-local outer-sheath strength/angular support} \\
		&\quad\rightarrow \text{proton energy/divergence regime}
	\end{aligned}
	\label{eq:full_mechanism_chain}
\end{equation}

Figure~\ref{fig:mechanism_robustness}(b) presents this sequence graphically by reading the quantitative regime contrasts collected in Fig.~\ref{fig:mechanism_robustness}(a) from optical buildup to the final proton response. The figure should therefore be read as a structured return to the independently reported values, not as an additional causal fit.

Each arrow is supported by an independently stored diagnostic, but the chain is intentionally regime-scale rather than particle-by-particle causal tracking. For the Representative case, the temporal ordering is consistent with a finite 50--100-fs accumulation/response interval between the early coherent electron/sheath organization and the later proton angular minimum; the available snapshots do not define a unique response time.

This interpretation refines our earlier fixed-geometry C-target picture~\cite{RezaeiPandari2026Arxiv}. The earlier comparison linked the higher cutoff primarily to enhanced electromagnetic coupling and a hotter energetic-electron population, while the curved geometry reorganized the surface-resolved and angular ion populations; it did not isolate a separate cutoff gain from geometric convergence alone. The present result shows that once the cavity geometry and laser focus are allowed to vary, efficient early optical coupling can increase both sheath strength and the angular domain of acceleration, improving energy while broadening the energetic population; more localized coupling and sheath support favor lower divergence at reduced cutoff. Persistent internal reflections alone do not determine that ordering.

This separation of energy and angular quality is consistent with earlier structured-target studies but uses a more explicit objective. Hemisphere--cone targets exhibit energy-selective focusing that changes between open and closed geometries~\cite{King2024PPCF}, whereas ultrathin-foil work reports an approximately $2^\circ$ half-angle component under a different divergence definition~\cite{Bin2013POP}. Neither supplies the centroid-referenced $E_k\geq5$~MeV RMS objective used here. Single-objective Bayesian PIC searches established data-efficient parameter selection~\cite{Dolier2022NJP}, while a recent double-layer multi-objective preprint retains peak energy, spectral purity, and in-window charge as separate objectives~\cite{Zhang2026Arxiv}. The present contribution is therefore the constrained Pareto resolution of cutoff energy against divergence for a curved cavity, together with matched-stage mechanism analysis, rather than a claim that all structured targets obey a universal energy--divergence law.

The principal numerical conclusions are likewise metric dependent. Grid and domain tests support the cutoff/divergence trends and the Representative cross-check, while PPC refinement leaves a larger residual sensitivity in integrated proton efficiency. The thin-wall mechanism survives the targeted \(\lambda_0/32\rightarrow\lambda_0/40\) analysis at the early strong-sheath stage, but the exact shell-integrated charge amplitudes are only resolution-checked. The late-time finite-box cutoff is not an asymptotic energy. These restrictions are retained rather than folded into a stronger convergence claim.

Several broader limitations remain. Each campaign design is represented by one production-resolution PIC realization, so the inferred Gaussian-process noise and the targeted grid/PPC tests do not constitute a repeated-seed estimate of stochastic PIC variability. The large separations among the three selected regimes and the numerical refinements support the reported ordering, but repeated-seed ensembles would be required to assign statistical uncertainty to individual Pareto points. The simulations are two-dimensional, and the reconstructed energy/divergence diagnostic neglects \(p_z\); the reported angular width is therefore a 2D Cartesian observable and should not be converted directly into a 3D cone angle. The model also cannot predict out-of-plane electromagnetic or particle leakage, full three-dimensional sheath topology, or absolute particle number and yield. The prescribed pre-plasma is shared across designs as a controlled modelling assumption rather than obtained from geometry-specific hydrodynamics, so robustness of the Pareto ordering to different experimental pre-plasmas remains untested. Electron transverse spreading, sheath curvature, and target-geometry effects can differ in three dimensions, so the numerical Pareto front and optimized parameter values are specific to the present simulation model rather than direct 3D experimental prescriptions. The intended contribution is a mechanism-informed design framework linking target geometry and focusing to optical coupling, electron redistribution, non-local sheath organization, and the proton energy--divergence trade-off; quantitative transfer to three-dimensional or experimental configurations requires dedicated validation. The optimized domain is finite, especially at the lower wall-thickness bound, so the Pareto front is a statement about the explicitly explored and validated design space rather than proof of a global physical optimum. Particle-history tracking of the terminal Max-Energy tail also remains beyond the present regime-scale mechanism analysis. These limitations define natural directions for future 3D and experimentally constrained optimization rather than weakening the present comparison among the tested regimes.


\section{Conclusions}
\label{sec:conclusion}

We used constrained multi-objective Bayesian optimization to map the energy--divergence trade-off of C-shaped targets in laser-driven proton acceleration. The final 82-design campaign contains 60 feasible evaluations and 15 constrained Pareto-optimal solutions. The Max-Energy endpoint reaches \(28.64~\mathrm{MeV}\) with \(52.51^\circ\) divergence and 13.90\% proton efficiency, while the Min-Divergence endpoint reaches \(11.47~\mathrm{MeV}\) with \(11.13^\circ\) divergence and 9.11\% efficiency. A Representative compromise reaches \(21.44~\mathrm{MeV}\), \(38.44^\circ\), and 12.90\%.

Matched-stage diagnostics show that these Pareto regimes are physically distinct. The Max-Energy design develops the strongest and broadest early outer sheath, whereas the lower-divergence regimes localize strong acceleration to narrower angular support. Electron kinetic-energy loading and near-interface charge separation reproduce the same regime ordering, while the charge-to-field relation remains non-local rather than pointwise. Optical diagnostics further show that geometry and focus establish different local field-buildup and early energy-partition conditions even when incident energies are nearly identical. The same data reject a simpler monotonic trapping-duration picture: propagation-corrected primary-reflection delays are nearly equal across the three cases, and the persistent secondary-reflection signature is strongest in the Representative design rather than in Max-Energy.

The resulting mechanism is a hierarchical design chain from target geometry and focusing to optical coupling, electron redistribution, near-interface charge-separation organization, outer-sheath strength/angular support, and finally the proton energy--divergence regime. The Representative case additionally exhibits a robust transient collimation of energetic protons around \(\tau\simeq100\) fs, consistent with a finite response interval following the earlier coherent electron/sheath stage. By contrast, the origin of the progressively collimated terminal high-energy tail in Max-Energy remains unresolved and is not folded into the global mechanism.

Targeted grid, domain, late-time, PPC, and thin-wall checks support the principal energy/angular trends and the early regime-scale charge/sheath interpretation. The cutoff remains a fixed-time rather than asymptotic observable, integrated proton efficiency retains residual PPC sensitivity, and shell-integrated thin-wall charge amplitudes are resolution-checked rather than formally converged. Within these limits, the study demonstrates that energy and angular quality can be treated as explicit competing design objectives and that the resulting Pareto regimes can be connected to a physically resolved laser--plasma mechanism rather than to geometry alone.

\section*{Funding}
This work was funded by the Iran National Science Foundation (INSF) under Project No.~4036338.

\section*{Acknowledgments}
The authors gratefully acknowledge the support of the Iran National Science Foundation (INSF).

\section*{Data and code availability}
The data and analysis code supporting the findings of this study are available from the corresponding author upon reasonable request.


\section*{AI-assisted manuscript preparation}
During manuscript preparation, the authors used OpenAI ChatGPT  for English-language editing, structural compression, consistency checking, literature-search organization, and assistance in drafting and synthesizing portions of the manuscript from author-provided simulation diagnostics. The tool did not generate primary simulation data, determine reported numerical values, or replace the authors' quantitative analysis. All numerical results originate from author-run Smilei simulations and author-controlled analysis workflows. The authors reviewed and verified the scientific interpretation, numerical values, code outputs, references, and final wording and take full responsibility for the content.

\section*{Competing interests}
The authors declare no competing interests.

\begingroup
\small
\bibliographystyle{unsrt}
\bibliography{HPLSE_refs,HPLSE_prior_work}
\endgroup

\end{document}